\documentclass{xpkas}

\def\year{2014} 
\def\month{May} 
\def\volume{00} 
\def\issue{0} 
\def\beginpage{1} 
\def\endpage{99} 
\def\received{September 30, 2014} 
\def\accepted{, 2014} 
\date{Received \received ; accepted \accepted}

\title{
Foreground of GRBs from \textit{AKARI} FIS data
}

\author[1]{L.~Viktor~T\'oth}
\author[2]{Yasuo~Doi}
\author[3,1]{Sarolta~Zahorecz}
\author[1,4]{M\'arton~\'Agas}
\author[1,4]{Lajos~G.~Bal\'azs}
\author[1]{Adrienn~Forr\'o}
\author[1]{Istvan~I.~R\'acz}
\author[5]{Zsolt~Bagoly}
\author[6]{Istv\'an~Horvath}
\author[1]{S\'andor~Pint\'er}

\affil[1]{Department of Astronomy, E\"otv\"os Lor\'and University Budapest; \email{lvtoth@astro.elte.hu}}
\affil[2]{University of Tokyo; \email{doi@ea.c.u-tokyo.ac.jp}}
\affil[3]{European Southern Observatory, Garching bei Munchen; \email{szahorec@eso.org}}
\affil[4]{Konkoly Observatory, Budapest; \email{balazs@konkoly.hu}}
\affil[5]{Department of Physics of Complex System, E\"otv\"os Lor\'and University Budapest}
\affil[6]{National Univ. of Public Service, Budapest}

\def\corrauthor{
L.V. T\'oth
}

\def\runningauthor{
L.V. T\'oth
}

\def\runningtitle{
GRB foreground from \textit{AKARI} FIS
}

\def\keywords{
infrared: galaxies, ISM -- gamma rays: bursts -- ISM: structure -- gamma-ray burst: individual: GRB060117 -- galaxies: individual: 2MFGC 16496
}

\def\abstracttext{
A significant number of the parameters of a gamma-ray burst (GRB) and its host galaxy are calculated from the afterglow. There are various methods obtaining extinction values for the necessary correction for galactic foreground. These are: galaxy counts, from HI 21 cm surveys, from spectroscopic measurements and colors of nearby Galactic stars, or using extinction maps calculated from infrared surveys towards the GRB.
We demonstrate that \textit{AKARI} Far-Infrared Surveyor sky surface brightness maps are useful uncovering the fine structure of the galactic foreground of GRBs. Galactic cirrus structures of a number of GRBs are calculated with a 2 arcminute resolution, and the results are compared to that of other methods.
}

\begin{document}
\pkashead 


\section{Introduction\label{sec:intro}}

It is always a challenge to accurately estimate the column density of the galactic foreground interstellar medium in the direction of extragalactic sources. It is also one of the important parameters when calculating the physical parameters of gamma-ray burst (GRB) host galaxies. We started an investigation of the infrared sky brightness towards GRBs using \textit{AKARI} Far-Infrared Surveyor (\textit{AKARI} FIS) of \cite{kawada2007} all-sky maps
\footnote{The sky maps can be retrieved from the data archive web site: \textit{http://www.ir.isas.jaxa.jp/ASTRO-F/Observation/}}.
\cite{doi2015}.
GRBs are the most energetic explosions in the Universe. A massive star undergoes core collapse, or a double neutron star or a neutron star and a black hole binary merges \cite{woosley2006}. X-ray and optical afterglows can outshine the brightest quasars. The redshift distribution of Swift GRBs shows that these objects may provide information up to high z values on: galaxy evolution, star formation history, intergalactic medium, see eg. \citet{gomboc2012}.
Most of the known physical parameters of the GRB and the GRB host galaxy are calculated from the afterglow. An estimate on the galactic foreground hydrogen column density towards the GRB is part of the calculations. It is based on galaxy counts and HI \citep{burstein1982}, HI surveys eg. the LAB survey \citet{kalberla2005}, extinction maps calculated from infrared surveys \citet{schlegel1998} and \citet{schlafly2011}, or from spectroscopic measurements and colors of nearby Galactic stars.

\section{Analysis of the \textit{AKARI} FIS All Sky Survey images \label{sec:2}}
\citet{doi2012}, and recently \citet{doi2015} have processed full sky images of the \textit{AKARI} FIS at 65$\mu$m, 90$\mu$m, 140$\mu$m and 160$\mu $m. The images achieve a detection limit of $< 10$\,MJysr$^{-1}$ with absolute and relative photometric accuracies of $< 20$\%. The spatial resolution of the survey is 1$'$. We substracted 30 by 30$\square '$ images centered on 283 GRBs with known redshifts used by \citet{horvath2014} in their analysis (see references therein). We selected 30 images for a test of foreground FIR emission, these were GRBs with associated FIR extragalactic sources \citet{toth2016} and GRBs associated to the large-scale Universal structure Hercules Corona Borealis Great Wall at a redshift of $z \approx 2$ by \citet{horvath2014}.

The color temperature maps of the large grain emission were estimated using the 90$\mu$m, 140$\mu$m and 160$\mu$m images. The maps were convolved to a 2$'$ resolution and, for each pixel, the spectral energy distribution (SED) was fitted with B$_\nu$(T$_{dust}$)$\nu^\beta$ with a fixed $\beta$=2.0 spectral index. The column densities averaged over a 2$'$ beam were calculated using the following equation with the intensity and temperature values from the SED fits:
\begin{equation} 
N(H)=\frac{2I_\nu}{B_\nu(T)\kappa\mu m_H}
\end{equation}
We used $\mu$=2.33 for the particle mass per hydrogen molecule and a dust opacity $\kappa$ obtained from the formula 0.1cm$^2$/g ($\nu$/1000 GHz)$^\beta$.

\section{FIR foreground\label{sec:3}}

\subsection{The foreground galaxy of GRB\,060117\label{sec:31}}
\begin{figure}[h]
\centering
\includegraphics[width=80mm]{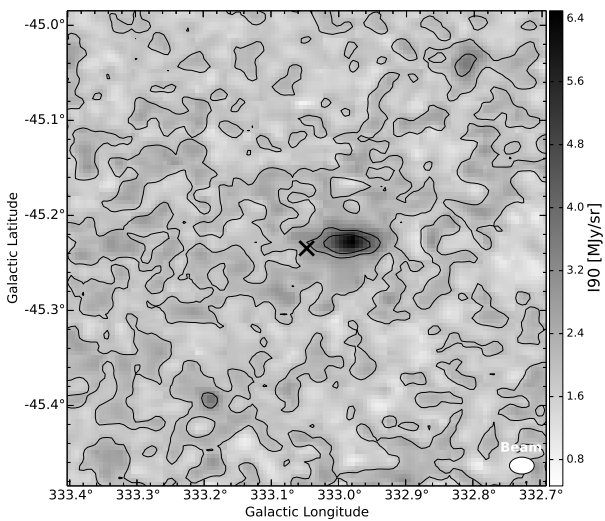}
\caption{\textit{AKARI} FIS 90\,$\mu$m image of GRB\,060117. Black cross indicates the GRB's position. The contour levels are at 2, 3.1 and 4\,MJysr$^{-1}$. The foreground galaxy west of the position of GRB\,060117 is 2MFGC\,16496. The \textit{AKARI} 90\,$\mu$m beam size is indicated in the lower right corner as a white ellipse.\label{fig:pkasfig1}}
\vspace{5mm} 
\end{figure}

ISM rich galaxies may have a size 2 times larger than their apparent optical size when measured from HI 21 or FIR data. We looked for GRBs with associated \textit{AKARI} galaxies from \citet{toth2016} and selected the 5 closest associations.
The 90 $\mu$m images have both a high spatial resolution $93\times 64 \square "$ and a high enough sensitivity to detect galaxies. One of the fields, the one centered on GRB\,0601175 showed a foreground galaxy 2-3 times more extended in FIR than it's NIR size of 0.8$'$ by 0.2$'$ \citet{skrutskie2006}. 
GRB\,060117 is a "long" type GRB with duration of 25\,s \citep{campana2006} at a photometric redshift of $z=0.98\pm0.24$ \citep{xiao2011}.
The bright FIR object is 2MFGC\,16496 a flat galaxy as appears in 2MASS images \citep{mitronova2004} at $z\approx0.0042$ \citep{jones2009}, i.e. it is clearly a foreground object. In optical and NIR images the foreground galaxy is relatively far from the GRB. Its galactic disk however is rather extended and apparently increases the foreground FIR sky brightness towards the GRB by approximately 1\,MJysr$^{-1}$. That emission by 2MFGC\,16496 may mislead us estimating the galactic FIR foreground, unless it is carefully subtracted. See Figure~\ref{fig:pkasfig1} for the \textit{AKARI} FIS 90\,$\mu$m image of the $30\times 30 \square '$ surroundings of GRB\,060117. The lowest contour is set at 3 times the standard deviation over the minimum surface brightness in the field.

\subsection{Structure of the Galactic foreground of GRBs in the Hercules Corona Borealis Great Wall\label{sec:32}}
\citet{horvath2013} discovered a concentration of $1.6<z<2.1$ GRBs in the Hercules-Corona Borealis region. Detailed statistical tests by \citet{horvath2014} indicate a significant clustering of those GRBs, that is also called as "the Hercules Corona Borealis Great Wall". This huge structure lies ten times farther away than the Sloan Great Wall \citep{gott2005}. The size of the structure defined by these GRBs is about 2000-3000\,Mpc, or more than six times the size of the Sloan Great Wall or more than twice the size of the Huge Large Quasar Group \citep{cloves2013}.

We investigated the structure of galactic foreground ISM of 24 GRBs all belong to the Hercules Corona Borealis Great Wall. A constant color temperature in the line of sight was estimated pixel-by-pixel using \textit{AKARI} FIS all sky survey 90, 140, 160\,$\mu$m images.  We assumed an emissivity of $\beta =2$. The distribution of the hydrogen column density $N($H$)$ was derived, as described in Section~\ref{sec:2}. The galactic foreground cirrus structures show a fluctuation on 3-4$'$ scale, sometimes with small chains of knots in the whole column density range.
In order to test the accuracy of our column density estimates based on lower angular resolution data, we calculated the averages $\bar{N}($H$)_6$ in a radius of 3$'$ and $\bar{N}($H$)_{30}$ for the $0.5 \times 0.5 \square '$ surroundings of the GRBs.
We compared the calculated column density averages with the central column density ($N($H$)_c$) value towards the GRB. A linear correlation was found for the 24 tested directions with a relatively large scatter. The linear correlation coefficients were 0.57 and 0.28 for the $\bar{N}($H$)_6$ vs. $N($H$)_c$ and the $N($H$)_{30}$ vs. $N($H$)_c$, respectively.
In as many as 50\% of the directions the $\bar{N}($H$)_6 - N($H$)_c$ difference was over $\pm$30\% of $N($H$)_c$ GRBs, and for 40\% the $\bar{N}($H$)_{30}$ average was more then 100\% off.

The \textit{Planck Space Telescope} \citep{tauber2010} observed the sky in 9 frequency bands covering 30 - 857\,GHz.  The \textit{Planck} images at 545 and 857\,GHz (550\,$\mu $m and 350\,$\mu $m respectively) have a spatial resolution of approximately 5$'$ \citep{planck2011}, similarly to IRIS \citep{miville2005}, the 100\,$\mu $m calibrated \textit{IRAS}   images.
We compared the \textit{AKARI} based column density estimates with estimates derived from IRIS 100$\,\mu m$, \textit{Planck} 857\,GHz and 575\,GHz images, and in general a good correlatio nwas found. A more detailed analysis including the use of the DustEM model \citep{compiegne2011} will be given elsewhere.

\subsection{Galactic foreground and the hydrogen column density of the GRB host galaxies\label{sec:33}}
We estimated the effect of the galactic foreground correction on the calculated hydrogen column density of the GRB host galaxy $N($H$)_{host}$. 
We downloaded spectra from the \textit{Swift-XRT} GRB Catalogue\footnote{\url{http://www.swift.ac.uk/xrt_live_cat/} maintained by the UK Swift Science Data Centre (UKSSDC)}, and analyzed those with Xspec\footnote{Xspec is part of the HEASOFT Software package of NASA's High Energy Astrophysics Science Archive Research Center (HEASARC), available at \url{http://heasarc.gsfc.nasa.gov/lheasoft/download.html}} \citep{arnaud1996}. The \textit{Swift-XRT} spectral data provided by the UKSSDC is calibrated and has the appropriate format for Xspec. We used exactly the same model as in the automatic analysis of the UKSSDC \citep{evans2009}. Each spectra was fitted with an absorbed power law with two absorbing components. The first component takes the Galactic foreground into consideration, and it is held fixed during a fit, while the second component gives the absorption due to the excess hydrogen column which is determined by the fitting. 

We selected test GRBs with a range of X-ray flux, at different galactic latitudes (that means varying $N($H$)_{foregr}$), and with a range of the initial values of $N($H$)_{host}$. We altered the $N($H$)_{foregr}$ foreground column density ($\pm 50$\%) and recalculated $N($H$)_{host}$. A 50\% increase or decrease of the assumed $N($H$)_{foregr}$ resulted in 15 to 35\% change of $N($H$)_{host}$. We consider that as a non-negligible difference.

\section{Conclusions}
Our tests indicate that a careful examination of the FIR foreground may in one hand reveal foreground FIR objects, on the other hand a high resolution mapping of the galactic cirrus foreground may significantly increase the accuracy of the estimation of foreground extinction. \textit{AKARI} FIS sky survey images are the proper data for that foreground analysis, that may serve as a basis for a recalculation of GRB host parameters.



\acknowledgments
This research is based on observations with AKARI, a JAXA project with the participation of ESA. This work made use of data supplied by the UK Swift Science Data Centre at the University of Leicester. This research was supported by OTKA grants NN111016 and K101393  and JSPS KAKENHI Grant 25247016.




\end{document}